\documentclass[11pt,a4,twoside]{article}
\usepackage[dvips,a4paper,left=2.5cm,right=2.5cm,top=2.5cm,bottom=2.5cm,headsep=1em]{geometry}
\usepackage{fancyhdr}
\usepackage{titlesec}
\usepackage[latin1]{inputenc}
\usepackage{amsmath,amsfonts,amssymb,array,amsthm}
\usepackage{rotating}
\usepackage[font={small}]{caption}
\usepackage{natbib}
\usepackage{lmodern,slantsc}
\usepackage{multirow}
\usepackage{chngcntr}
\usepackage{placeins}
\usepackage{caption}
\usepackage{enumitem}
\usepackage[breaklinks,
colorlinks=true, linkcolor=blue, citecolor=black, urlcolor=blue,
pdfborder={0 0 0}, pdfpagelabels]{hyperref}
\newcommand{\hshift}{\hspace{0.875mm}}

\def\v0{\boldsymbol{0}}

\newlength{\FigureHeight}
\newlength{\FigureHeightHalf}

\numberwithin{equation}{section}
\titleformat{\section}
{\large\bfseries}{\thetitle.}{0.5em}{}
\titleformat{\subsection}
{\normalfont\bfseries}{\thetitle.}{0.5em}{}
\titleformat{\subsubsection}
{\normalfont\itshape}{\normalfont\thetitle.}{0.5em}{}
\begin{document}

\title{\vspace{-3.75em} Symmetries and turbulence modeling. A critical examination}
\author{George Khujadze$\,^1$\thanks{Email address for correspondence:
george.khujadze@uni-siegen.de}$\:\,$ \& Michael Frewer$\,^2$ \\ \\
\small $^1$ Chair of Fluid Mechanics, Universit\"at Siegen, 57068
Siegen, Germany\\
\small $^2$ Heidelberg, Germany}
\date{{\small\today}}
\clearpage \maketitle \thispagestyle{empty}

\vspace{-2em}\begin{abstract}

\noindent
The recent study by \cite*{Klingenberg20} proposes a new strategy for modeling turbulence in general. A proof-of-concept is presented therein for the particular flow configuration of a spatially evolving turbulent planar jet flow, coming to the conclusion that their model can generate scaling laws which go beyond the classical ones. Our comment, however, shows that their proof-of-concept is flawed and that their newly proposed scaling laws do {\it not} go beyond any classical solutions. Hence, their argument of having established a new and more advanced turbulence model cannot be confirmed. The problem is already rooted in the modeling strategy itself, in that a nonphysical statistical scaling symmetry gets implemented. Breaking this symmetry will restore the internal consistency and will turn all self-similar solutions back to the classical ones.
To note is that their model also includes a~second nonphysical symmetry. One of the authors already acknowledged this fact~for turbulent jet flow in a formerly published Corrigendum \citep*{Sadeghi20}.\linebreak[4]
However, the Corrigendum is not cited and so the reader is not made aware that their method has fundamental problems that lead to inconsistencies and conflicting results. Instead, the\linebreak[4] very same nonphysical symmetry gets published again. Yet, this unscientific behaviour is not corrected, but repeated and continued in the subsequent and further misleading publication \cite*{Klingenberg22}, which is examined in this update in the appendix.

\vspace{0.5em}\noindent{\footnotesize{\bf Keywords:} {\it Statistical Physics, Turbulence Modeling, Lie Groups, Symmetry Analysis, Jet~Flows}}
\end{abstract}

\section{Introduction\label{Sec1}}

To validate their newly developed modeling ansatz in \cite{Klingenberg20}, which essentially resulted into what they called `minimal' turbulence model, Eqs.(73-76), the authors give a proof-of-concept in Sec.IV.E, along with App.D1. By choosing the particular flow configuration of a spatially evolving turbulent planar jet flow, they come to the conclusion that their model can generate scaling laws, as~Eq.(80), which go beyond the classical ones and therefore can describe effects which classical scaling laws aren't capable of. For example, by stating on p.12:

\vspace{0.25em}
\textit{``~...~the new model equations behave more like the exact equations by also allowing for the generalized self-similar scaling given by (80), rendering the new model more general and, thus, reliable than the existing ones."},

\vspace{0.25em}\noindent
or, on p.17:

\vspace{0.25em}
\textit{``~...~the form derived here is more general and, thus, holds promise to be valid for a wider range of the flow. In particular, form (80), assuming that $A_2$ is slightly larger than 1, can describe the frequently observed behavior that the Reynolds stresses (and higher correlations) take longer to achieve self-preservation (in the classical sense) than the mean velocity."}

\vspace{0.25em}\noindent
But, unfortunately, the authors did not examine their new scaling laws Eqs.(79-80) carefully enough. For example, let's first consider the original and {\it non}-modeled statistical system of balance equations, Eq.(5), for the axial and transverse momentum of a (spatially evolving) planar turbulent jet flow in their exact and complete form
\begin{equation}
\overline{U}_j\frac{\partial\overline{U}_i}{\partial x_j}=-\frac{\partial\overline{P}}{\partial x_i}+\nu\frac{\partial^2 \overline{U}_i}{\partial x^2_j}
-\frac{\partial R^{(0)}_{ij}}{\partial x_j},\quad i,j=1,2,
\label{200302:1404}
\end{equation}
together with the continuity equation, Eq.(4),
\begin{equation}
\frac{\partial \overline{U}_i}{\partial x_i}=0, \quad i=1,2.
\label{200727:2047}
\end{equation}
Equation \eqref{200302:1404} for the axial momentum $(i=1)$ gives rise to the following integral invariant, the globally constant momentum flux (see e.g. \cite{Coles17})
\begin{equation}
J=\int_{-\infty}^\infty \Big(\overline{U}_1^2+R^{(0)}_{11}-\nu\partial_{x_1}\overline{U}_1+\overline{P}-P_\infty\Big)dx_2,
\end{equation}
while no corresponding invariant emerges from the transverse momentum equation $(i=2)$, due to the specified boundary conditions defining this flow \citep{Pope00,Davidson04,Coles17}
\begin{equation}
\left.
\begin{aligned}
&\overline{U}_1\big\vert_{x_2=\pm\infty}=0,\quad \partial_{x_2}\overline{U}_1\big\vert_{x_2=\pm\infty}=0, \\[0.5em]
&\overline{U}_2\big\vert_{x_2=\pm\infty}=V_0\ll 1,\quad \partial_{x_2}\overline{U}_2\big\vert_{x_2=\pm\infty}=0,\\[0.5em]
&R^{(0)}_{ij}\big\vert_{x_2=\pm\infty}=0,\quad \overline{P}\big\vert_{x_2=\pm\infty}=P_\infty=\text{const}.
\end{aligned}
~~~~~\right\}
\end{equation}
The condition $V_0\neq 0$ is needed to describe or implement the process of entrainment of outer fluid into the jet flow in order to feed the jet, in particular when the viscous dragging forces are negligibly small \citep{Davidson04}. Ultimately, $V_0\neq 0$ is the reason that the transverse momentum flux is not globally conserved.

Now, if we treat the planar turbulent jet flow as laid out above in its statistically exact form without any approximations, we are dealing with two {\it intrinsic} global parameters, $J$ and~$\nu$, which allow us to define a natural global length and time scale of the flow:
\begin{equation}
L=\frac{\nu^2}{J},\qquad T=\frac{\nu^3}{J^2}.
\label{200727:1942}
\end{equation}
These two global scales are necessary if we want to define certain local scales of the flow, say a local length scale~$\ell$ and a local velocity scale $u$. Since the aim herein is to study the self-similar region(s) of a spatially evolving planar flow, we search for local scales either dependent on the axial coordinate $x_1$ or the transverse coordinate $x_2$ \citep{Coles17}. As in \cite{Klingenberg20},\linebreak[4] we choose the former dependence:
\begin{equation}
\ell=\ell(x_1),\qquad u=u(x_1).
\end{equation}
The global scales \eqref{200727:1942} can then be used to express them in their dimensionally correct form
\begin{equation}
\ell(x_1)=L\cdot\tilde{\ell}\big(x_1/L\big),\quad\;
u(x_1)=(L/T)\cdot\tilde{u}\big(x_1/L\big),
\label{200727:2026}
\end{equation}
where $\tilde{\ell}$ and $\tilde{u}$ are now dimensionless functions, where it goes without saying that these functions are unknowns of the system and therefore not known beforehand.

Now, to search for self-similar regime(s) in a solution, we make the general ansatz of reducing the number of coordinates, in going here for a spatially evolving flow from two spatial variables $(x_1,x_2)$ to only one spatial variable~$\tilde{x}$. In particular, we make the ansatz of a dimensionless self-similar variable \citep{Coles17}:
\begin{equation}
(x_1,x_2)\:\rightarrow\: \tilde{x}=\frac{x_2}{\ell(x_1)},
\label{200727:2345}
\end{equation}
which therefore, since $\tilde{x}$ is dimensionless, leads to an ansatz of dimensionless field functions
\begin{equation}
\overline{U}_i(x_1,x_2)\:\rightarrow\: \tilde{U}_i(\tilde{x}),\quad
R_{ij}^{(0)}(x_1,x_2)\:\rightarrow\: \tilde{R}_{ij}^{(0)}(\tilde{x}),
\end{equation}
but which, with the local velocity scale \eqref{200727:2026}, can be scaled into their dimensionally correct form
\begin{equation}
\frac{\overline{U}_i(x_1,x_2)}{u(x_1)}=\tilde{U}_i(\tilde{x}),\quad
\frac{R_{ij}^{(0)}(x_1,x_2)}{u^2(x_1)}=\tilde{R}_{ij}^{(0)}(\tilde{x}).
\label{200727:2040}
\end{equation}
Whether this ansatz for a specifically modeled local scale $\ell$ and $u$ is valid or not within a certain flow regime can only be decided by evaluating relations \eqref{200727:2040} through experimental or simulation data. If this is the case, then such a self-similarity is called a {\it self-similarity of the first~kind} \citep{Barenblatt96}, since ansatz \eqref{200727:2040}, with its self-similar variable \eqref{200727:2345}, was established by dimensional analysis. In clear contrast to a solution if it shows a {\it self-similarity of the second kind} which cannot be established by dimensional analysis alone and therefore cannot be detected by the simple ansatz \eqref{200727:2040}. Such self-similarities are characterized by additionally having to solve a nonlinear eigenvalue problem in order to determine the self-similar variable \citep{Barenblatt96}.
Nevertheless, when the ansatz for the self-similar variable is based on the particular form~\eqref{200727:2345}, the above procedure is the only correct way to find the corresponding self-similar regime
in a spatially evolving planar turbulent jet flow when considering its statistically exact and non-modeled form~\eqref{200302:1404}-\eqref{200727:2047}.

In practice, however, the exact equations \eqref{200302:1404}-\eqref{200727:2047} get reduced to inviscid equations, with the argument that for a high enough Reynolds number ($Re\sim 1/\nu \gg 1$) the viscous terms are negligibly small in relation to all other terms. Although this reduction constitutes an approximation of \eqref{200302:1404}-\eqref{200727:2047}, it is the following resulting system of (Euler) equations
\begin{equation}
\left.
\begin{aligned}
&\frac{\partial \overline{U}_i}{\partial x_i}=0, \quad i=1,2,\\[0.5em]
&\overline{U}_j\frac{\partial\overline{U}_i}{\partial x_j}=-\frac{\partial\overline{P}}{\partial x_i}
-\frac{\partial R^{(0)}_{ij}}{\partial x_j},\quad i,j=1,2,
\end{aligned}
~~~~~\right\}
\label{200727:2108}
\end{equation}
which are standardly called in the literature, and thus also in \cite{Klingenberg20}, as the exact statistical equations (exact up to viscous stresses) of a spatially evolving planar turbulent jet flow.
However, regarding the {\it intrinsic} global and local scales, the inviscid system \eqref{200727:2108} is fundamentally different from the viscous one \eqref{200302:1404}-\eqref{200727:2047}. Instead of two global scales, $J$~and $\nu$, we now only have a single global scale to work with, the globally constant momentum flux $J$, which for the invsicid case takes the reduced form \citep{Coles17}
\begin{equation}
J=\int_{-\infty}^\infty \Big(\overline{U}_1^2+R^{(0)}_{11}+\overline{P}-P_\infty\Big)dx_2.
\end{equation}
The problem now is that we are missing one global scale in order to construct a global length and time scale so that local scales can get dimensionalized appropriately. This problem, however, can only be solved by letting  the global scales change into local ones, where naturally we will choose here $L$ to be $x_1$, which then will define $T$. Thus instead of \eqref{200727:1942}, the global scales for the inviscid case are local:
\begin{equation}
L=x_1,\qquad T=\frac{x_1^{3/2}}{J^{1/2}}.
\label{200728:1910}
\end{equation}
This result then turns the two local scales \eqref{200727:2026}~to
\begin{equation}
\ell(x_1)=x_1\cdot\tilde{\ell}(1),\quad\;
u(x_1)=(J/x_1)^{1/2}\cdot\tilde{u}(1),
\label{200728:1909}
\end{equation}
and therefore result \eqref{200727:2345} and \eqref{200727:2040} finally~to
\begin{equation}
\frac{\overline{U}_i(x_1,x_2)}{(J/x_1)^{1/2}}=\tilde{U}_i(\tilde{x}),\quad
\frac{R_{ij}^{(0)}(x_1,x_2)}{J/x_1}=\tilde{R}_{ij}^{(0)}(\tilde{x}),
\label{200727:2338}
\end{equation}
with the uniquely determined self-similar variable
\begin{equation}
\tilde{x}=\frac{x_2}{x_1}.
\label{200728:1833}
\end{equation}
\pagebreak[4]

\noindent
The dimensionless constants $\tilde{\ell}(1)$, $\tilde{u}(1)$ in \eqref{200728:1909} have been absorbed into the dimensionless functions $\tilde{U}_i$
and~$\tilde{R}_{ij}^{(0)}$.
Hence, for the invisicid case, result \eqref{200727:2338} is the only physically consistent ansatz to search for self-similar solutions in a spatially evolving planar turbulent jet flow, if, and this is important, if the self-similar variable is set to be of the form \eqref{200728:1833}. Any other ansatz which makes use of the same self-similar variable \eqref{200728:1833}, but which is not structurally and functionally equivalent to \eqref{200727:2338}, will be inconsistent in one way or the other, as the one proposed by Eq.(80) in \cite{Klingenberg20}.

\section{Refuting the demonstrating example in \cite{Klingenberg20}\label{Sec2}}

While Eq.(79) corresponds to the self-similar velocity result obtained in \eqref{200727:2338} and thus is correct, it is Eq.(80) for the Reynolds stresses which is flawed. Although Eq.(80)
\begin{equation}
R_{ij}^{(0)}=\tilde{H}_{ij}(\tilde{x})x_1^{-A_2}-\tilde{U}_i(\tilde{x})\tilde{U}_j(\tilde{x})x_1^{-1},\quad \tilde{x}=x_2/x_1,
\label{200728:1901}
\end{equation}
was obtained with the Lie-group symmetry method, which itself allows to investigate the similarity solutions of the first and second kind systematically, it results in a scale $x_1^{-A_2}$ which, for $A_2\neq 1$,
has no physical origin of the flow considered. Since the similarity solution~\eqref{200728:1901} is defined with respect to the very same self-similar variable as~\eqref{200728:1833}, it therefore {\it cannot} be identified as a self-similarity of the second kind, simply because \eqref{200728:1833} is the result of a dimensional scaling \eqref{200728:1910}-\eqref{200728:1909} and thus, by construction, associated to a similarity of the first kind --- that \eqref{200728:1901} indeed constitutes a similarity of the {\it first kind} is also independently supported by the fact that \eqref{200728:1901} rests on three {\it global scaling} symmetries, as clearly expressed in the derivation of \eqref{200728:1901} through Eqs.(D3-D4) in \cite{Klingenberg20}. But, as already explained before, a physically consistent self-similar solution of the {\it first kind} with respect to the dimensionless self-similar variable $\tilde{x}=x_2/x_1$ \eqref{200728:1833}, is and can only be given by \eqref{200727:2338}
\begin{equation}
R_{ij}^{(0)}=J\cdot\tilde{R}_{ij}^{(0)}(\tilde{x})x_1^{-1},
\label{200728:2020}
\end{equation}
where the dimensionless function can further be represented or decomposed into the $H$-framework of \cite{Klingenberg20}:
\begin{equation}
\tilde{R}_{ij}^{(0)}(\tilde{x})=\tilde{H}_{ij}(\tilde{x})-\tilde{U}_i(\tilde{x})\tilde{U}_j(\tilde{x}).
\label{200728:2021}
\end{equation}
Hence, when comparing \eqref{200728:2020}-\eqref{200728:2021} with \eqref{200728:1901}, we see that the latter solution can only be physically consistent if~$A_2=1$. This clearly indicates that a scaling based on $A_2\neq 1$ is nothing but a pure mathematical artefact, which, as will be discussed later in more detail at the end of this section, is rooted in the fact that one of the three scalings to derive \eqref{200728:1901} is not a true symmetry but only a superficial equivalence transformation resulting from the statistically unclosed equations \eqref{200727:2108}, which, by closer inspection, even turns out to be a nonphysical equivalence transformation.

That consistency of \eqref{200728:1901} can only be restored if $A_2=1$, can further be independently demonstrated by considering the free-shear boundary-layer approximation of the governing system \eqref{200727:2108},
in which the two momentum equations are reduced to a single balance equation in the axial momentum only \citep{Bradbury65,Rotta72,Lumley94},
\begin{equation}
\left.
\begin{aligned}
\hspace{-0.2cm}&\frac{\partial \overline{U}_i}{\partial x_i}=0,\;\; i=1,2,\\[0.5em]
\hspace{-0.2cm}&\overline{U}_j\frac{\partial\overline{U}_1}{\partial x_j}=-\frac{\partial R^{(0)}_{12}}{\partial x_2}
-\frac{\partial\big(R^{(0)}_{11}-R^{(0)}_{22}\big)}{\partial x_1},\;\; j=1,2,
\end{aligned}
~~\right\}
\label{200729:1109}
\end{equation}
if we, as before, assume the jet to issue into a quiescent background fluid having constant pressure and being sufficiently far from any solid boundaries, but now using the additional fact that in a planar jet the transverse component of the mean inertial term is negligibly small compared to\linebreak[4]
\pagebreak[4]

\noindent
the streamwise component. System~\eqref{200729:1109} can then be further approximated~to \citep{Davidson04,Pope00,Lumley94,Schlichting17}
\begin{equation}
\left.
\begin{aligned}
&\frac{\partial \overline{U}_i}{\partial x_i}=0, \quad i=1,2,\\[0.5em]
&\overline{U}_1\frac{\partial\overline{U}_1}{\partial x_1}+\overline{U}_2\frac{\partial\overline{U}_1}{\partial x_2}=-\frac{\partial R^{(0)}_{12}}{\partial x_2},
\end{aligned}
~~~~~\right\}
\label{200302:1405}
\end{equation}
by recognizing that also the axial gradients in the Reynolds stresses are much smaller than the transverse gradients, i.e.,
\begin{equation}
\frac{\partial R^{(0)}_{ij}}{\partial x_1}\ll \frac{\partial R^{(0)}_{ij}}{\partial x_2}, \quad \forall i,j.
\label{200304:1114}
\end{equation}
Equation \eqref{200302:1405} is a well-established approximation throughout literature,
and is clearly validated both by experiment and numerical simulation within the approximation it assumes. Sure, the whole approximation itself can be criticized \citep{Coles17} as not being accurate enough in certain specific regions of the jet flow,
but this is outside the scope of our comment.
The issue here is whether the newly proposed scaling \eqref{200728:1901} can appear
and thus be a possible solution of
the self-similar flow regime of~\eqref{200729:1109}, or \eqref{200302:1405}, when the self-similar variable is based on the form $\tilde{x}=x_2/x_1$~\eqref{200728:1833}, especially since equation \eqref{200302:1405} is unanimously accepted to be accurate enough to correctly describe that particular regime of the flow \citep{Davidson04,Pope00,Lumley94,Coles17,Rotta72,Bradbury65,Schlichting17}.

Now, if Eqs.(79-80) in \cite{Klingenberg20} are the correct and true new scaling laws, then they should balance not only the exact equations \eqref{200727:2108}, which as a mathematical artefact they indeed do,\footnote{That the reduction of the Lie-group symmetry analyzed PDE-system \eqref{200727:2108} to an ODE-system is indeed a mathematical artefact when scaling it according to Eqs.(79-80) in \cite{Klingenberg20},
will be explained and discussed in detail at the end of Sec.$\,$\ref{Sec2}.} but they should also balance the approximate PDE-system \eqref{200302:1405} and reduce it to an approximate ODE-system in the invariant variable $\tilde{x}=x_2/x_1$. But such a balance can only occur if~$A_2=1$, as can be clearly seen when scaling \eqref{200302:1405} according to Eqs.(79-80):
\begin{equation}
\left.
\begin{aligned}
0&=\frac{\partial \overline{U}_i}{\partial x_i}
=-\frac{\tilde{U}_1^\prime}{x_1^{3/2}}\tilde{x}-\frac{\frac{1}{2}\tilde{U}_1-\tilde{U}_2^\prime}{x_1^{3/2}},\\[0.5em]
0&=\overline{U}_1\frac{\partial\overline{U}_1}{\partial x_1}+\overline{U}_2\frac{\partial\overline{U}_1}{\partial x_2}+\frac{\partial R^{(0)}_{12}}{\partial x_2}\\[0.5em]
&=-\frac{\tilde{U}_1\tilde{U}_1^\prime}{x_1^2}\tilde{x}
-\frac{\frac{1}{2}\tilde{U}_1^2+\tilde{U}_1\tilde{U}_2^\prime}{x_1^2}
+\frac{\tilde{H}_{12}^{\prime}}{x_1^{A_2+1}}\\[0.5em]
&=-\frac{2\tilde{U}_1\tilde{U}_2^\prime}{x_1^2}
+\frac{\tilde{H}_{12}^{\prime}}{x_1^{A_2+1}},
\end{aligned}
~~~~~\right\}
\label{200729:1403}
\end{equation}
where prime denotes the derivative with respect to the invariant variable $\tilde{x}$. The same problem arises for system~\eqref{200729:1109}. Any deviation from the classical value $A_2=1$ will break the self-similarity in a non-approximative manner, simply because no self-similar ODE with respect to the single variable $\tilde{x}$ can be constructed, neither for \eqref{200302:1405} nor for \eqref{200729:1109}.

Hence, in contrast as to what is claimed in \cite{Klingenberg20}, their newly proposed scaling laws Eqs.(79-80) do not go beyond any classical solutions, irrespective of whether we consider an exact balance equation, as \eqref{200727:2108}, or an approximate one, as \eqref{200729:1109} or \eqref{200302:1405}. For the exact\linebreak[4] balance equation this is shown through result \eqref{200728:2020}-\eqref{200728:2021}, while for the approximate equations it is representatively shown through \eqref{200729:1403}, that no other value except the classical value $A_2=1$ can lead to consistent self-similar solutions when based on the self-similar variable $\tilde{x}=x_2/x_1$.

\newgeometry{left=2.5cm,right=2.5cm,top=2.5cm,bottom=2.0cm,headsep=1em}

Now, what further does the result $A_2=1$ imply for the overall conclusion in \cite{Klingenberg20}?
Well, since according to definition Eqs.(D9-D10)\footnote{Please note that Eq.(D8) in \cite{Klingenberg20}, and therefore also Eq.(D9), has a misprint in the scaling exponent.
Instead of `$2a_{\text{Sc},I}$'~in the denominator, it should only be `$a_{\text{Sc},I}$'.}
\begin{equation}
A_2=-\frac{2a_{\text{Sc},I}-2a_{\text{Sc},I\!I}+a_{\text{Sc,stat}}}{a_{\text{Sc},I}},
\end{equation}
and the fact that
\begin{equation}
\frac{a_{\text{Sc},I}-a_{\text{Sc},I\!I}+a_{\text{Sc,stat}}}{a_{\text{Sc},I}}=-\frac{1}{2},
\end{equation}
due to the constant spreading rate Eq.(D7) of the jet, the restriction $A_2=1$ forces the group parameter of one particular scaling symmetry to be zero:
\begin{equation}
a_{\text{Sc,stat}}=0.
\label{200729:1455}
\end{equation}
A result which comes as no surprise, because this `symmetry' is well-known to be nonphysical. Result \eqref{200729:1455} is just one further proof that joins a long list of proofs that already have been given for various other turbulent flow configurations, e.g., as for channel flow with and without wall transpiration \citep{Frewer14.2,Frewer16.2}, or for temporally evolving jet flow \citep{Frewer18.2}. In particular, this `statistical scaling symmetry', explicitly given in \cite{Klingenberg20} by Eq.(41), has been proven nonphysical even in its most general form within different statistical frameworks and without specifying any particular flow configuration \citep{Frewer15.1,Frewer16.1,Frewer16.3,Frewer17}.

The key aspect of this statistical scaling Eq.(41) is that it's not a true symmetry transformation but only some arbitrary equivalence transformation\footnote{More precisely, Eq.(41) is a physically {\it non}-realizable equivalence transformation, meaning that although Eq.(41) leaves the originally unclosed statistical equations invariant, the transformed statistical fields
cannot be physically realized \citep{Frewer14.2,Frewer15.1,Frewer16.1,Frewer17}. In other words, according to the transformation rule of Eq.(41), if the non-transformed fields $\overline{U}_i$, $\overline{P}$, $H_{ij}$, etc.,
define a statistical solution set of the deterministic Navier-Stokes equations, then the transformed set of statistical fields $\overline{U}_i^*$, $\overline{P}^*$, $H^*_{ij}$, etc.,
do {\it not} form statistical solutions anymore, which can emerge or which can be generated from the deterministic and thus closed Navier-Stokes equations. They are thus not realizable and therefore, ultimately, nonphysical.}
which emerges because the underlying statistical equations \eqref{200727:2108} are simply unclosed. In clear contrast to the other two scaling symmetries Eq.(24) and Eq.(25), driven by the group parameters $a_{\text{Sc},I}$ and $a_{\text{Sc},I\!I}$, which are true symmetry transformations since they emerge from the corresponding symmetries Eqs.(11-12) of the statistically non-averaged and thus closed Euler equations.
The difference between these two invariant transformations is that a symmetry transformation maps a solution of a specific (closed) equation to a new solution
of the same equation, while an equivalence transform acts in a weaker sense in that it only maps an (unclosed) equation to a new (unclosed) equation of the same class --- for more details, see e.g. \cite{Frewer14.2} and the references therein, as well as the detailed remark R4 on p.13 in \cite{Frewer18.1}.

In other words, it is the all-embracing closure problem of turbulence that generates this analytical arbitrariness, which too, with the method of Lie-group transformations, cannot be bypassed. Since the statistical equations of turbulence are unclosed, so is their set of invariant Lie-group transformations. Unclosed equations, as those considered in \cite{Klingenberg20} to determine their new `symmetries',\footnote{Note that the crucial `symmetries' Eq.(41-44) in \cite{Klingenberg20} are determined from the unclosed equations Eqs.(17-19), and not from the modeled equations,
because this is exactly what the aim of \cite{Klingenberg20} is,
namely to extract from the unclosed equations some invariant transformations to then use these to close those same equations again. An endeavour which is comparable to pull oneself up by one's own bootstraps.} inevitably lead to infinite dimensional Lie algebras, which means that (nearly) any invariant transformation can be generated, and thus also (nearly) any desirable scaling law. Ultimately one has an
infinite set of invariant possibilities to choose from when performing a full and correct Lie-group symmetry analysis for unclosed equations, as shown e.g. in \cite{Frewer14.1,Frewer16.4,Frewer18.2}. A crucial information which is not shared with the reader in \cite{Klingenberg20}.

\restoregeometry

\newgeometry{left=2.5cm,right=2.5cm,top=2.5cm,bottom=2.5cm,headsep=1em}

Hence, the Lie-group symmetry method in turbulence is {\it not} free of any assumptions. It~is an ad-hoc method too, not in the same but in a similar way as the classical self-similarity method: Instead of using an {\it a priori} set of scales, the Lie-group method has to make use of an {\it a priori} set of symmetries, namely to select the correct relevant symmetries from an infinite (unclosed) set. In other words, the particular selection of the additionally chosen symmetries Eqs.(41-44) in \cite{Klingenberg20} is an assumption and {\it not} a result that comes from theory, as they misleadingly try to convey. Because, as just referenced before, when correctly performing a complete and systematic Lie-group symmetry analysis on the considered set of unclosed equations \eqref{200727:2108}, including its infinite hierarchy of all unclosed higher order correlation equations as presented in their study \cite{Oberlack10}, one gets an infinite set of functionally independent invariances, and not only those few reported in \cite{Oberlack10}\linebreak[4] and presented again through Eqs.(34-44) in \cite{Klingenberg20} --- to note is that all `new' invariances in \cite{Oberlack10}, or equally in \cite{Rosteck13}, were obtained only through heuristics and a trial-and-error ansatz, and {\it not} through a complete and systematic Lie-group analysis, which would have given an unclosed set of invariances and thus an overall different conclusion, namely that the Lie-group method alone, like any other analytical method, cannot bypass the closure problem of turbulence.

Another concerning issue not mentioned in \cite{Klingenberg20} is the fact that due to the arbitrariness involved when making a particular choice from an infinite (unclosed) set of possible symmetries, there is a high chance that one will select a nonphysical symmetry which is not reflected by experiment or numerical simulation. This clearly is the case in \cite{Klingenberg20}, not only for the chosen scaling symmetry Eq.(41), but also for the arbitrarily chosen set of translation symmetries Eqs.(42-44). Although knowing better that this set of symmetries is nonphysical and evidently not supported by experiment or simulation, as clearly acknowledged in \cite{Sadeghi20}, it nevertheless gets published and used again in their current study \cite{Klingenberg20} to model turbulent~flows.

Important to note here is that the statistical translation symmetry Eqs.(42-44) in \cite{Klingenberg20} is exactly the same symmetry as in \cite{Sadeghi20,Sadeghi18}, and not any variant thereof. Specifically, Eqs.(42-44) only appears as a sub-group of a more general symmetry in \cite{Sadeghi20,Sadeghi18}. The group parameters
$a_{\text{Tr,stat,$I$,$i$}}$, $a_{\text{Tr,stat,$I\!I$,$ij$}}$, $a_{\text{Tr,stat,$I\!I\!I$,$i$}}$
are named therein as $a_{U_i}$, $a_{H_{ij}}$, $a_{PU_i}$, respectively.
As already discussed in \cite{Sadeghi18}, the group parameter $a_{U_i}$ has to be put to zero since it is not compatible with a physical constraint, particularly with the conserved mass flow rate of a jet, while $a_{H_{ij}}$ had to be forced to zero in \cite{Sadeghi20} as the authors finally realized that the statistical translation symmetry as a whole is nonphysical, meaning that finally also $a_{PU_i}$ has to be put to zero in order to be overall consistent. Hence, the relevance and impact of the statistical translation symmetry Eqs.(42-44) for a turbulent jet flow is void. A crucial information which is not revealed to the reader\footnote{What's additionally disturbing here is that instead of citing the JFM-Corrigendum \cite{Sadeghi20}, the flawed original paper \cite{Sadeghi18} gets cited as Ref.(33), however not self-critical, but in
the straight opposite way, presented as a paper that has no faults at all. In particular it is praised as an overall successful paper,
not only once but several times at a stage where the nonphysical translation symmetry Eqs.(42-44) has already been introduced on p.6.} in \cite{Klingenberg20}, thus misleading the reader into the wrong belief that the chosen symmetry Eqs.(42-44) is a valid and physical symmetry for {\it all} possible flow configurations, simply because no configuration in \cite{Klingenberg20} gets singled out by the authors when establishing their generic turbulence model Eqs.(73-76).

But explicitly excluding the jet-flow configuration from their analysis and thus from their newly proposed generic model Eqs.(73-76), would have
required an explicit explanation by the authors to counter the question:
{\it Why to exclude from a generic turbulence model such a basic canonical flow as jet flow?} It's clear that exclusion of such a basic flow configuration will give the impression that something is fundamentally wrong with this model, which leads us to the next section.

\restoregeometry

\newgeometry{left=2.5cm,right=2.5cm,top=2.5cm,bottom=2.5cm,headsep=1em}

\section{Strong indications that the new modeling strategy by Klingenberg et al. is methodologically flawed\label{Sec3}}

The root problem in \cite{Klingenberg20} is the implementation of the two nonphysical statistical symmetries Eq.(41) and Eqs.(42-44).
Although the authors recognize already early on that there is a natural barrier Eq.(68) that prevents them from implementing these two symmetries,
they nevertheless force this implementation by artificially pushing up the number of variables, which, obviously, is not a very advantageous and promising procedure as a modeler.\footnote{As a modeler one usually prefers to use Occam's razor as a basis, in that unnecessarily complex models should not be preferred to simpler ones.} In~fact, this blind focus on symmetries, especially in turbulence modeling as exercised in \cite{Klingenberg20}, limits creativity considerably.\footnote{In this context it is worth reading \cite{Hossenfelder18}, to realize that the arguments presented therein may also apply to turbulence modeling.} Freely adapted from Mies van der Rohe, it's here valid to say:

\vspace{-0.25em}
\noindent\hfill {\it ``Symmetry is the tool of the unimaginative"}\hfill\phantom{.}

\subsection*{The points of criticism of the new modeling principle}

\noindent
\textbf{1.} Despite the fact that the two symmetries Eqn.(41) and Eqs.(42-44) are unphysical {\it per se},\linebreak[4] in that they both violate the classical principle of
cause and effect \citep{Frewer14.2,Frewer15.1,Frewer16.1,Frewer17,Frewer18.2}, their key modeling-argument in itself is circular:

{\it Two new `symmetries', Eqn.(41) \& Eqs.(42-44), are extracted from unclosed equations to then use them in order to find improved closing constraints for those same~(!) equations again.}
This effort is comparable to pull oneself up by one's own bootstraps.

\vspace{0.75em}
\textbf{1.1.} As already mentioned in Sec.$\,$\ref{Sec2}, the set of possible statistical symmetries is infinite, simply because from the outset the statistical equations themselves are unclosed. These equations, as well as their symmetries, need to be modeled empirically. To take any of such (unmodeled) symmetries, in order to exactly model those same equations again from which they originated, is of no physical value, since the modeling information in this situation has to come from the outside (experimental and/or numerical data) and cannot come from the inside, in particular not when the intended modeling-element, namely the set of symmetries, is itself unclosed and need to be modeled for its own use. To note in this regard is that the set of symmetries in the functional Hopf-framework is infinite and therefore unclosed too, alone already by the fact that the linear superposition principle exists and applies in this particular framework.

\textbf{1.2.} With modeling of symmetries, we mean to find which is the correct symmetry to take~(if~any) when the pool of symmetries to choose from is infinite. The aim is to avoid the high risk of choosing nonphysical symmetries. On the contrary, a statistical symmetry can be regarded as correct or as approximately correct in the modeling sense if it's more or less consistent to the statistical data considered. However, for the two new statistical symmetries Eq.(41) and Eqs.(42-44), this is definitely not the case, as they both fail to be consistent with numerical data already for simple canonical cases as turbulent channel flow \citep{Frewer14.2,Frewer16.2}, or turbulent jet flow \citep{Frewer18.2,Sadeghi20}. Excluding these two nonphysical symmetries from the modeling process clearly shows that the matching to the DNS data improves by several orders of magnitude, which can be well explained from the fact that both symmetries Eq.(41)~and Eqs.(42-44) violate the classical principle of causality (for a review, see e.g. \cite{Frewer15.4}).

\textbf{1.3.} Their motivation to choose from an infinite pool of possibilities exactly only those two symmetries, Eq.(41) and Eqs.(42-44), is based on the following overall false claim:

\vspace{0.0em}
\textit{``Generally speaking, the statistical symmetries (41)-(44) encode important principles of turbulent statistics, namely, intermittency and non-Gaussianity." [p.12].}

\pagebreak[4]
\restoregeometry

\newgeometry{left=2.5cm,right=2.5cm,top=2.5cm,bottom=2.0cm,headsep=1em}

\vspace{0.0em}\noindent
This alleged connection to intermittency and non-Gaussianity has been clearly refuted several times by now within different statistical frameworks \citep{Frewer15.1,Frewer15.2,Frewer16.3}. Also
from a pure phenomenological viewpoint, it's abundantly clear that intermittency is a symmetry-breaking phenomenon and not a symmetry-existing or symmetry-preserving one \citep{Frisch95}. Even if we would wrongly assume this to be the case, intermittency is definitely not described or featured by any {\it global} scaling symmetry, particularly not by the one given by Eq.(41), which actually just mimics the standard scaling of a linear system when applied to any unclosed nonlinear system, namely in the way that it simply identifies all nonlinear terms just as error terms which only need to be corrected for by exploiting the unclosed terms. Obviously, such a symmetry is only a mathematical artefact of the unclosed system itself and therefore indeed nonphysical, independent of the fact, of course, that this symmetry also violates the classical principle of cause and effect.

In fact, spatial-global symmetries, i.e., symmetries with constant group parameters, as Eq.(41), are not able to describe intermittent phenomena; at most, these can only be described by local space-dependent symmetries whose group parameters depend on the space coordinates. In~order to describe intermittent phenomena, regardless of whether they are on a small or large scale, multifractal scaling approaches are necessary \citep{Frisch95}. A global and universal scaling ansatz as presented in \cite{Klingenberg20} is therefore definitely not sufficient.

\vspace{1em}\noindent
\textbf{2.} From a pure mathematical viewpoint, the introduction and necessity of the new velocity field~$\hat{U}_i$ and pressure field $\hat{P}$ is clear. But not from a physical viewpoint! What exactly is the physical meaning, for example, of~$\hat{U}_i$?

\vspace{0.75em}
\textbf{2.1.} Sec.IV.D of \cite{Klingenberg20} is aimed to give some physical meaning to these new additional fields. But their explanation fails.

To interpret these variables as some kind of update- or correction-variables of the mean velocity and pressure fields, $\overline{U}_i$ and $\overline{P}$, as the authors are trying to~do, is misleading. The reason is that the dynamical equations Eqs.(73-74) of the new variables $\hat{U}_i$ and~$\hat{P}$ are fully decoupled from the variables $\overline{U}_i$ and $\overline{P}$ and, hence, cannot update or correct these in any iterative way. Instead, the mean dynamics of $(\overline{U}_i,\overline{P},H^{(0)}_{ij},\dotsc)$ is {\it globally} defined by these new modeling variables~$(\hat{U}_i,\hat{P})$, however, in such a peculiar way that they are solutions of~the very same (inviscid) Navier-Stokes equations again, Eqs.(73-74), which in \cite{Klingenberg20} were actually intended to be modeled. Hence, they obtain the full (inviscid) Navier-Stokes equations as the modeled equations
again, which means that their modeling strategy is even based on a further, second circular argument:

\vspace{0.25em}
{\it Equations get modeled by equations which are exactly the same~(!) as those being modeled.}

\vspace{0.25em}\noindent
We are thus thrown back to the start, because since the model-defining equations Eqs.(73-74) are the same exact equations as the (inviscid) Navier-Stokes equations, they again suffer from the same problems which originally were intended to be modeled, namely the instability problem, which, depending on the input of initial and boundary conditions, leads to chaotic and turbulent solutions, which again need to be modeled statistically. Hence, as a result of this circular model, the natural question arises on which length and time scales these new fields $(\hat{U}_i,\hat{P})$ operate or should operate~on? An answer or any discussion is not given in \cite{Klingenberg20}, but if done, will result in a negative outcome \citep{Frewer19}. In fact, as explained and discussed
in \cite{Frewer19}, all these new fields are physically fictitious and do not exist.

\textbf{2.2.} Finally the proof-of-concept in Sec.IV.E, to demonstrate how successful the new model might be, also fails. This was clearly shown in Sec.$\,$\ref{Sec2}. Already this failure is indication enough that their modeling approach is fundamentally flawed in the way how compatibility is forced for symmetries which are nonphysical, resulting then in an artificial model involving new variables for which no physical interpretation can be given.

\vspace{1em}\noindent
\textbf{3.} In all said, it should be clear that we do not criticize the method of Lie-groups itself, which is a very useful mathematical tool indeed, when only applied to the right problems.

\restoregeometry

\section{Further points for correction in \cite{Klingenberg20}\label{Sec4}}

This section lists several false and unfounded statements that have been made in \cite{Klingenberg20} and briefly explains why a correction is necessary in each case:

\vspace{1em}
\textit{``This has led to symmetry (42) frequently being used in turbulence modeling due to its similarity with the Galilean transformation." [p.6]}

\vspace{0.25em}\noindent
This claim comes without a reference. In fact, nowhere in the literature a study can be found that substantiates this claim. We would like to point out that this is not the first time that this particular symmetry (42) has been mischaracterized by one of the present authors. M.~Oberlack falsely identified and characterized symmetry (42) in previous publications \citep{Oberlack14.1,Oberlack14.2} as Kraichnan's random Galilean invariance group, in order to give (42) some physical meaning. Upon our comments \citep{Frewer15.1,Frewer15.2}, M.~Oberlack formally acknowledged this in his published reply \citep{Oberlack15} as a mistake, but downplayed it as being \textit{``only a matter of wording and has nothing to do with the mathematical or physical content of our paper." [\cite{Oberlack15}, p.4]}

\vspace{1em}
\textit{``Any model mainly concerned with stationary turbulence which was meant to make use of the Galilean group, hence, essentially included the statistical symmetry (42)." [p.6]}

\vspace{0.25em}\noindent
Also this claim comes without any reference. In fact this claim is wrong as no model can be found which is based on such assumptions, simply because there is no link between the Galilean group
and the statistical symmetry~(42). The aforementioned `similarity' for stationary flows is misleading and should not be used to justify symmetry~(42). The simple reason is that the Galilean symmetry is based on a coordinate transformation while~(42) is~not! \citep{Frewer15.1}

\vspace{1em}
\textit{``In fact, the symmetry (42) is essential in any turbulence model to mimic the well-known logarithmic law of the wall, as is discussed in the work of Oberlack Ref.(29)." [p.6]}

\vspace{0.25em}\noindent
This claim has been clearly refuted in \cite{Frewer14.1,Frewer14.2,Frewer16.2}.

\vspace{1em}
\textit{``~...~(42)-(44) extended by the higher moments depict the non-Gaussianity of turbulence Ref.(34)." [p.6]}

\vspace{0.25em}\noindent
This claim is wrong and seriously misleading. Ref.(34) clearly does not support this claim, simply because the probability density function (PDF) formulation of the translation symmetry Eqs.(42-44) is not compatible with all the physical PDF-constraints with which this formulation inherently comes along with. In fact, the technically and methodologically flawed study Ref.(34) has been refuted in its entirety \citep{Frewer15.1,Frewer15.2,Frewer16.3,Frewer17}.

\vspace{1em}
\textit{``However, it guarantees that invariant solutions such as the log law or an extended jet flow scaling law such as that found in the work of Ref.(33) can be generated using any combination of the symmetries considered." [p.8]}

\vspace{0.25em}\noindent
This claim is incorrect as not any combination of the considered symmetries guarantees the generation of physical scaling laws. In Ref.(33) the statistical translation symmetry Eqs.(42-44) needs to be excluded from this process, as was already acknowledged before by one of the authors in \cite{Sadeghi20}.

\vspace{1em}
\textit{``The expectation that a model developed within the presented framework will, indeed, have an improved predictive quality partly relies on the great significance of the statistical symmetries for predicting the scaling of turbulent flows, as discussed in Refs.(23,25,30-32) and Ref.(33)." [p.11]}

\vspace{0.25em}\noindent
This claim is wrong and misleading as in Ref.(33) no scaling prediction based on statistical\linebreak[4] 
symmetries\hfill was\hfill achieved.\hfill In\hfill Ref.(33)\hfill the\hfill statistical\hfill translation\hfill symmetry\hfill is\hfill nonphysical,\hfill as\hfill

\newgeometry{left=2.5cm,right=2.5cm,top=2.5cm,bottom=1.70cm,headsep=1em}

\noindent acknowledged
in \cite{Sadeghi20} and thus had to be put overall to zero, while the second statistical symmetry, the scaling symmetry, already had been put to zero in Ref.(33) since from the outset it was not compatible with a physical constraint: \textit{``~...~it immediately follows that the scaling symmetry implied by $a_1$ ... break[s] and give[s] $a_1=0$." [\cite{Sadeghi18}, p.244]}. Hence, none of the statistical symmetries that originally have been considered in Ref.(33) are left to make any scaling predictions!

Note again that the nonphysical statistical translation and scaling symmetry in Ref.(33) correspond exactly to Eqs.(42-44) and Eq.(41) in \cite{Klingenberg20}, respectively.
Also note that all their other references mentioned in the above quotation are misleading too: Refs.(23,$\,$25) and Ref.(32) suffer from the very same problems as Ref.(33), while the two older references Refs.(30,$\,$31) do not deal with statistical symmetries at all, since they were not `invented' before~2010 with the publication of~Ref.(25).

To summarize this section, we note again that although the authors of \cite{Klingenberg20} knew of the JFM-Corrigendum \citep{Sadeghi20} as a crucial publication that has a significant impact on the interpretation and conclusion of their currently published turbulence model, the JFM-Corrigendum is neither cited nor mentioned anywhere in the current study. Instead, the flawed original study \cite{Sadeghi18} gets cited as Ref.(33) to misleadingly promote the false narrative that every advanced turbulence model should possess the two symmetries Eq.(41) and Eqs.(42-44).

\appendix

\section{Review of \cite{Klingenberg22}\label{SecA}}

In\hshift this\hshift updated\hshift second\hshift version\hshift (June\hshift 2022),\hshift we\hshift provide\hshift a\hshift review\hshift of\hshift\cite{Klingenberg22}\linebreak[4]
to show that it follows the same line of reasoning as their previous study \cite{Klingenberg20} we refuted herein.
Nothing is presented that might challenge or invalidate our critique. Instead, their new result, when properly examined, only confirms our findings, assertions and conclusions.
Thus, all the sections above from the first version (Secs.$\,$\ref{Sec1}-\ref{Sec4}) are unchanged in this updated version.

The key aspect to recognize in \cite{Klingenberg22} is that although their study is based on an inconsistent and methodologically flawed modeling approach, it nevertheless arrives at a reasonable and comparable result for turbulent plane jet flow, as shown in Figs.1-2. Why is that so? The answer is simple: because their particular result shown in Figs.1-2 does not reflect the inconsistent modeling structure they have developed in the sections before. The study therefore ends in a wrong conclusion. Two features of their final result make this clear:
\begin{itemize}
\item[{\bf 1.}]
The new scaling symmetry (36) of the closed model equations (68)-(74) in \cite{Klingenberg22} is not related in any form to the nonphysical scaling symmetry (18) of the unclosed statistical equations (3)-(4). They are two fundamentally different symmetries\linebreak[4] and therefore it is wrong to assign (36) the same group parameter ($a_{\text{Sc,stat}}$) as~(18). If~done~so, as in the paper, it immediately leads to the following contradiction in the transformations:\\
While (36) transforms the turbulent kinetic energy simply as $k^*=e^{a_3}k$, the under\-lying and defining invariance~(18) transforms it completely different and unnaturally to
$k^*=e^{a_3}k+\frac{1}{2}(e^{a_3}-e^{2a_3})(\overline{U}_1^{\hspace{0.25mm}2}+\overline{U}_2^{\hspace{0.25mm}2}+\overline{U}_3^{\hspace{0.25mm}2})$,
and are identical only if all mean velocities $\overline{U}_i$ are zero, but which for turbulent plane jet flow is certainly not the case. Note that the latter transformation is based on the Reynolds decomposition $H_{ij}=R_{ij}+\overline{U}_i\overline{U}_j$, and $a_3$ is the alias for the lengthy group parameter $a_{\text{Sc,stat}}$.\\
And therefore, since the model equations (68)-(74) are invariant only under the newly constructed third scaling symmetry (36) (next to (34) and (35)), and not under (18), a~connection to the nonphysical symmetry (18) is not given, as incorrectly claimed in \cite{Klingenberg22}. That (18) is nonphysical and thus non-realizable has been rigorously proven in the literature several times by now (see e.g. \cite{Frewer22.1} and the references therein).
The question of whether (36) is physical or not does not arise here, since this symmetry was specifically designed to keep invariant an arbitrary but closed (2+1)-equation model based on the Boussinesq approximation. For (36), the question is rather whether it is relevant or not. At this stage of their paper, however, no conclusion can be drawn, simply because the particular solution shown in Figs.1-2 is not based on this symmetry (36). This brings us to the next point, where instead of looking at the model equations (68)-(74) and their invariance, as in this first point, we now look at the particular invariant solution that this system produces.

\item[{\bf 2.}]
The invariant solution shown in Figs.1-2 is not based on the new scaling symmetry (36). It is based on only the two classical Euler scaling symmetries (34) and (35). The reason is that the solution in Figs.1-2 emerges from the symmetry breaking integral constraint~(77), both for $\hat{U}_1$ and $\overline{U}_1$, which forces the group constants to reduce to $a_2=3a_1/2$ 
and $a_3=0$, with the aliases $a_1=a_{\text{Sc,I}}$ of (34), $a_2=a_{\text{Sc,II}}$ of (35) and $a_3$ as before, $a_3=a_{\text{Sc,stat}}$ of (36).\\
In other words, the integral constraint (77) breaks the third scaling symmetry (36) to $a_3=0$ and therefore does not take part in the invariant solution of Figs.1-2. A result which already has been found in \cite{Sadeghi18} (see p.244, where the integral constraint (2.4) breaks the third scaling symmetry of the temporally evolving plane jet~to~$a_1=0$).\\
For completeness, the concrete steps taken to obtain this symmetry-breaking result\linebreak[4] $a_2=3a_1/2$ and $a_3=0$ is first to generate the invariant functions of $\overline{U}_1$ and $\hat{U}_1$ from the symmetries (34)-(36), which are $\overline{U}_1=x^{1-a2/a1+a3/a1}\tilde{u}(y/x)$ and $\hat{U}_1=x^{1-a2/a1}\tilde{\hat{u}}(y/x)$, and then to demand invariance of the globally constant momentum flux (77) for both fields, exactly as it was done in the paper to yield their result (75).\\
Note that the paper has several sign typos: In (35) the scaling exponents for $k$, $\varepsilon$, $\hat{\varepsilon}$ must be negative, in (70)-(71) the right-hand sides must be positive, and in (72)-(74) all first terms on the right-hand side, i.e. the $C_\mu$-terms, must be positive too.
\end{itemize}

\noindent
{\bf Conclusion:}  Taking both these features into account, namely first that the simulated model equations (68)-(74) are not invariant under (18), which \cite{Klingenberg22} also incorrectly links to intermittency, and second that the new third scaling symmetry (36) is broken ($a_{\text{Sc,stat=0}}$) in the invariant solution shown in Figs.1-2, thus renders their central conclusion that {\it ``the intermittent edge of the jet is captured very well, which could be linked to the fulfillment of the statistical scaling symmetry (18)"} incorrect.

Hence, what is shown by the green lines in Figs.1-2 is just the classical invariance solution pressed into a new and more complicated set of equations. That's all. Their original aim to construct a system of model equations that is invariant under (18) is not achieved. And the reason for this failure is clear: (18) is simply an unrealizable invariance.

\subsection{Further points for correction in \cite{Klingenberg22}\label{SecA1}}

Besides their wrong conclusion that symmetry (18) leads or could lead to better results, their study is also afflicted with other problems and errors.

\vspace{0.75em}\noindent
\textbf{1.}
Not only the scaling (18) but also the translation symmetry (19-20) is nonphysical and thus non-realizable. In particular for jet flow, this has been explicitly demonstrated in the forced Corrigendum \cite{Sadeghi20}. Yet, this publication is again neither mentioned nor cited. Instead, only the flawed original paper gets cited as Ref.[6] in terms of a successful study again that shows {\it ``excellent agreement with wide ranges of experimental and numerical data"}~[p.4]. A~remarkable attitude that was already criticized two years ago. By now it's clear that an inconvenient fact is being withheld here from editors, referees and readers. To consistently ignore this negative result and to still sell Ref.[6] as a successful study already resorts to scientific misconduct.

\vspace{0.75em}\noindent
\textbf{2.}
When calibrating the newly constructed model for channel and pipe flows to the core-region deficit law on p.9, it is claimed that {\it ``the modified versions of both models are far more flexible, allowing for arbitrary values of $\sigma_1$"}, whereas the classical $k$--$\varepsilon$ model does not, predicting the wrong exponent $\sigma_1=1/2$. This is not true. Both the modified as well as the classical models allow for arbitrary values of $\sigma_1$. The mistake lies in analyzing the model equations globally, instead of using the local equations defining the core region. Because, if one inserts the scaling ansatz (59) into the model equations when they are localized around the central region of the flow, then both the modified as well as the classical models allow for arbitrary values of $\sigma_1$. Hence, the modified models show no advantage here.

Note that the modified and classical models both predict a wrong scaling of the turbulent kinetic energy in the core region. They both do not predict the correct value $n_k=2$. The reason that $n_k=2$ and no other value results from a trivial fact explained in \cite{Frewer22.1}, or in more detail in \cite{Frewer22.2}.

Also note that in the referred-to paper, cited as Ref.[7], the scaling ansatz for the turbulent kinetic energy in the core region of channel and pipe flow will result to an expression of the form $k=C_kx_2^{n_k}+F(\overline{U}_1)$, where $F$ is a fixed non-zero quadratic function in the mean velocity~$\overline{U}_1$ resulting from the Reynolds-decomposition of the instantaneous approach, while in (59) the ansatz is with $F(\overline{U}_1)=0$, which is fundamentally different. Which ansatz is correct now?

\vspace{0.75em}\noindent
\textbf{3.}
Question to the authors: Why does \cite{Klingenberg22} not follow up on the promising invariant ``solution" (80) from the earlier work \cite{Klingenberg20} and compared against experimental data? Is it because it is nonphysical as proven herein in Sec.$\,$\ref{Sec2} and therefore cannot be matched to data?

\vspace{0.75em}\noindent
\textbf{4.}
All modeling variables generated in the cited papers Refs.[24-27] have a clear physical origin, either through averaging or filtering of the Navier-Stokes equations. In clear contrast of course to the new modeling variables $\hat{U}_i$ and $\hat{\varepsilon}$ in their paper, which have no origin and appear out of nowhere just to forcibly fulfill an invariant setting of two nonphysical symmetries (18) and (19-20). Therefore, a connection or similarity to the method of selecting modeling variables in Refs.[24-27] is not given to justify the current approach.

Also, to justify the new modeling variables with the way $k$ and $\varepsilon$ are introduced in the classical $k$--$\varepsilon$ model, namely to say that {\it ``the link between the model variables $k$ and $\varepsilon$ and the mean velocity is not based on physical arguments, but on heuristics"} [p.6], is misleading. Of course, the relationship between the modeling variables in the classical $k$--$\varepsilon$ model is heuristic, but the variables themselves, $k$ and $\varepsilon$, have a clear physical origin and can be physically measured. This, obviously, is not the case for the new modeling variables and therefore are fundamentally different to the classical ones.

To also say that for the classical $\varepsilon$-equation {\it ``most modelers would agree that attempting to base it on the exact equation that can be derived for the dissipation would lead to an inferior overall model performance (e.g., Refs.13 and 28)"} [p.6] is a freely invented statement nowhere documented. Instead, in Ref.13 cited, the modeler Wilcox only says that modeling the exact dissipation equation is not any more rigorous than modeling a new equation based on dimensional-analysis arguments [p.89], but not that it leads to an inferior model performance. He also points out that it's the variables chosen that play the decisive role in modeling, because {\bf ``The physics is in the choice of variables"} [p.84].

A contradiction in their current paper is also to say on p.12 that the turbulent kinetic energy~$k$ is not an end in itself, but only a means to an end in order to predict the mean velocity as accurately as possible, while on p.9 exactly the opposite is done: The equation for $k$ is explicitly used to calibrate against experimental data to infer $C_\mu=0.09$.

\vspace{0.75em}\noindent
\textbf{5.}
Finally a remark to the numerical solution: The impression that the green line in Fig.1 fits the experimental data better seems to be an arbitrary result determined by the user and not uniquely given by the boundary condition at $\eta\rightarrow\infty$, as may be seen in Fig.2(a), where the field does not go to zero smoothly but experiences a jump close to $\eta=0.25$. It's not clear whether such an extra condition of zero values was imposed here just to achieve the desired faster (steeper) convergence to zero for the mean streamwise velocity field in Fig.1. What is also not clear is whether the originally stated boundary-value problem (BVP) for the new model equations (68)-(74) leads to a unique solution at all.

\bibliographystyle{jfm}
\bibliography{BibData}

\end{document}